\documentclass[prl,twocolumn,aps,footinbib,superscriptaddress]{revtex4}
\usepackage{amssymb}

\usepackage{amssymb}

\usepackage{graphicx}
\usepackage{amsmath}
\usepackage{times}
\usepackage{subfigure}
\usepackage{color}

\newcommand{\be}{\begin{equation}}
\newcommand{\ee}{\end{equation}}
\newcommand{\tr}{\mbox{Tr}}

\newcommand{\ket}[1]{\ensuremath{| #1 \rangle}}

\definecolor{blue}{rgb}{0,0,0.8}

\begin{document}

\title{Phase transitions, entanglement and quantum noise interferometry in cold atoms}

\author{Florian Mintert }
\affiliation{ Department of Physics, Harvard University, 17 Oxford Street, Cambridge Massachusetts, USA} \affiliation{
Physikalisches Institut, Albert-Ludwigs Universit\"at Freiburg, Hermann-Herder-Str. 3, Freiburg, Germany} \affiliation{
Institute for Theoretical Atomic, Molecular and Optical Physics, Harvard University, Cambridge, Massachusetts 02138,
USA}
\author{Ana Maria Rey}
\affiliation{ Institute for Theoretical Atomic, Molecular and Optical Physics, Harvard University, Cambridge,
Massachusetts 02138, USA}
\author{Indubala I. Satija}
\affiliation{ National Institute of Standards and Technology, Gaithersburg, Maryland 20899, USA} \affiliation{
Department of Physics, George Mason University, Fairfax, Virginia 22030, USA}
\author{Charles W. Clark}
\affiliation{ Joint Quantum Institute, National Institute of Standards and Technology and University of Maryland,
Gaithersburg, Maryland 20899, USA}

\begin{abstract}
We show that entanglement monotones can characterize the pronounced enhancement of entanglement at a quantum phase
transition if they are sensitive to long-range {\it high order} correlations. These monotones are found to develop a
sharp peak at the critical point and to exhibit universal scaling. We demonstrate that similar features are shared by
noise correlations and verify that these experimentally accessible quantities indeed encode entanglement information
and probe separability.
\end{abstract}
 \maketitle

Quantum entanglement represents one of the most fascinating features of quantum theory and has emerged as an important
resource in quantum information science \cite{Nielsen}. As entanglement represents a unique form of correlations that
do not occur in classical systems, the investigation of connections between entanglement and quantum phase transitions
is an emerging field of research which promises to complement the understanding of critical phenomena in condensed
matter physics and quantum field theory.

Iconic examples of exactly solvable models exhibiting quantum phase transitions, such as the Ising-like  spin chain,
are becoming  paradigms for investigating   the  suitability of  proposed  entanglement measures, e.g. the concurrence
and entanglement entropy \cite{Amico}, to quantify   the entangling resources of a system close to its quantum critical
point. Among the various entanglement measures the concurrence  has  \cite{Amico}  gained particular attention   in
view of its   universal scaling \cite{osterloh,Osborne} close to the phase transition. However,  the regular bipartite
concurrence has the drawback that  even at the critical point where spin-spin correlations extend over a long range (as
the correlation length is diverging for an infinite system) only the next and next-to nearest neighbor concurrences are
non zero. Moreover, it is not the concurrence that peaks at the transition but its first derivative.

The expected enhancement of entanglement at the phase transition has been observed with the help of generalizations of
the bipartite concurrence \cite{Oliveira} or the entanglement entropy $S_{\ell}$ between a block of $\ell$ consecutive
spins and the rest of the chain \cite{Latorre,Vidal}. However, there is no systematic understanding of why some
entanglement measures capture this enhancement whereas other fail to do so. In the present letter we provide a general
explanation for the different abilities of various entanglement measures to describe quantum phase transitions.

With the help of a sequence of entanglement measures
we illustrate the importance of {\it higher order } correlations in encoding long range correlations near quantum phase transition.
For this purpose we characterize the entanglement properties of a chain of interacting spins with different
entanglement monotones and study  how correlations are established over different ranges. We see that there is a
qualitatively different behavior of long-range correlations as opposed to their short-range analogues, as well as
between two- and many-body entanglement.

Whereas entanglement measures have become common tools for theoretical investigations, they are significantly less
popular in the experimental community since they are typically extremely arduous if not practically impossible to
measure. For the verification of theoretical findings and for the  investigation of theoretically intractable systems
it is therefore necessary to have efficient {\em experimental} means to probe entanglement properties. Among the
various physical systems exhibiting quantum phase transitions, cold atoms loaded in optical lattices offer considerable
advantages over more traditional  set-ups since they allow to engineer various Hamiltonians with high precision and to
include or remove noise and randomness in a controlled way. In these systems many-body correlations are extractable
from  absorption images of the atomic cloud after its release from the trap \cite{Altman,Foelling,Porto}. However such images
typically do not provide sufficient information to recover a proper entanglement measure. Here we give a general
prescription of how to extract the pronounced enhancement of entanglement at a phase transition from the limited
accessible experimental data.

We consider the ground state of the one dimensional spin-$1/2$ anisotropic  XY model in a transversal magnetic field
characterized by the Hamiltonian
\begin{equation}
\hat{H}=-\frac{J}{2}\sum_{j=1}^N (1+\gamma)\sigma_j^x\sigma_{j+1}^x+(1-\gamma)\sigma_j^y \sigma_{j+1}^y+h\sigma_j^z\ ,
\label{ham}
\end{equation}
where $J$ is the nearest neighbor coupling constant, $h$ is a transverse magnetic field, $\gamma$ is the anisotropy
parameter, $0 \leq \gamma \leq 1$, $\sigma_j^\alpha$ are Pauli matrices ($\alpha=x,y,z$) and periodic boundary
conditions are assumed throughout. For $0 < \gamma \leq 1$, $\hat H$ belongs to the Ising universality class and
exhibits a  quantum phase transition from a paramagnetic to a ferromagnetic phase when $\lambda=h/J$ takes its critical
value $\lambda_c=1$ \cite{Sachdev}. The XY model is  exactly solvable, and  any correlation function can be expressed
in terms of T\"{o}plitz-like determinants after a Jordan-Wigner transformation that maps spin operator into fermionic
operators\cite{Lieb, Rey}. The evaluation of {\it all}  four-point  correlations  is  however more cumbersome  than
the commonly used prescription for computing  two-point functions, and details about the procedure will be provided
elsewhere\cite{inpreparation}.

A severe problem for the investigations of entanglement in many-body systems is that such entanglement often {\em
cannot} be discovered in terms of few-body correlations. Thus, a few-body state that is obtained via a partial trace
over a many-body state often shows only weak or no entanglement even if the underlying many-body state is highly
entangled \cite{osterloh,Osborne}. Therefore, we will not consider entanglement properties of reduced few-body states,
but only investigate properties of the ground state of the entire many-body system. Instead of focusing on quantifying
the entanglement  remaining in  a subsystem of a few spins after tracing over the others, we focus on the entanglement
shared between these spins and the rest.  The latter does encode information about the entanglement of the pure
many-body state.

There is an enormous zoo of entanglement monotones and measures to quantify the entanglement of many-body systems
\cite{review_horo}, and there is no generally accepted choice that elevates one above the rest. For bipartite systems,
both entanglement entropy and concurrence play an important role, since they are the first monotones that could be
evaluated by purely algebraic means for mixed states \cite{concurrence1a,concurrence2}. Entanglement entropy is defined
in terms of the degree of mixing of the reduced density matrix $\varrho_r$ of either of the two subsystems:
$S(\Psi)=-\tr\varrho_r\log\varrho_r$, and the most frequently used definition of concurrence is
$c(\Psi)=|\langle\Psi^\ast|\sigma_y\otimes\sigma_y|\Psi\rangle|$, where $\sigma_y$ is the second Pauli matrix,
$\langle\Psi^\ast|$ is the complex conjugate of $\langle\Psi|$, that is the transpose of $|\Psi\rangle$, with the
conjugation/transposition performed in the eigenbasis of the third Pauli matrix $\sigma_z$. For a system of two
spin-1/2 objects concurrence and entanglement entropy are equivalent as one is a bijective function of the other.
However, for the case of multipartite systems that we consider here, this is no longer the case; in particular, there
is no unique generalization to many-body systems for either of these measures, and our choices are taken in accordance
with the currently investigated situation without any claim to be a canonical choice.

\begin{figure}
\centering \subfigure[]{\includegraphics[width=1.6in]{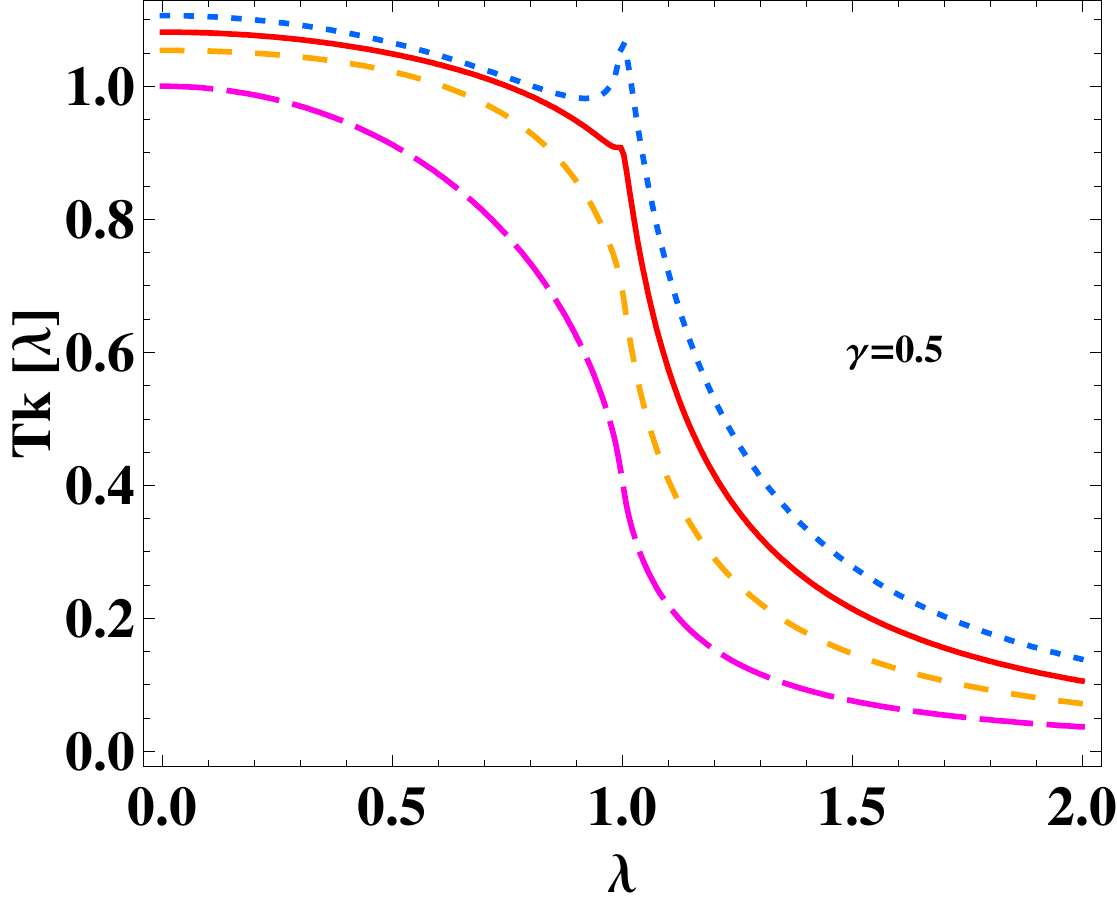}}\quad \subfigure[]{\includegraphics[width=1.6in]{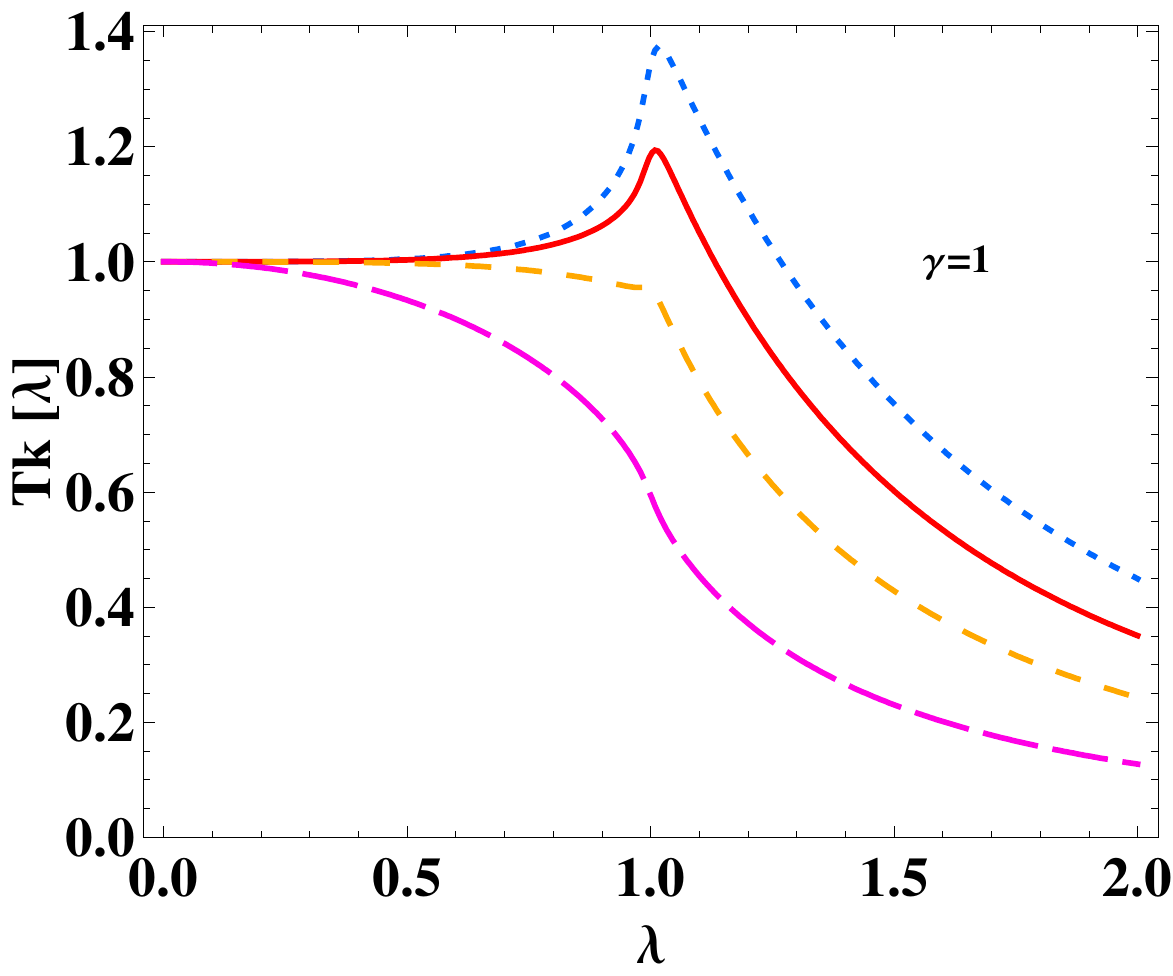}}
\caption{(color online) Various multipartite variations of the tangle (squared concurrence) $T_1$ long-dashed (pink),
$T_2$ dashed (yellow), $T_3$ solid(red), and $T_4$ dotted (blue) as function of $\lambda$ for $\gamma=1$ (right panel)
and $\gamma=0.5$ (left panel). $T_1$, and $T_2$ do not show a distinct peak around the phase transition. $T_k$ are
obtained by numerical computation of T\"{o}plitz determinants for a system of size $N=175$.} \label{fig1}
\end{figure}

In the case of entanglement entropy we will divide the spin chain into two blocks of not necessarily consecutive spins
and consider the entanglement entropy between those blocks. In the case of concurrence we will use generalizations
based on the popular redefinition $c(\Psi)=\sqrt{2(1-\tr\varrho_r^2)}$ \cite{rungta} of concurrence: we will focus on
the discrete set of generalized tangles (squared concurrence) \cite{marek}: \be
T_k=2-\frac{2}{D_k}\sum_{\nu\in\kappa_k}\tr\varrho_\nu^2 \ee where the $\varrho_\nu$ are reduced density matrices of
$1$ to $k$ spins, and the summation is performed over all reduced density matrices of up to $k$ spins. Formally, this
means that $\kappa_N$ contains all $1$ to $k$ touples of pairwise different elements labeling the individual sites. The
constants $D_k=\sum_{i=0}^{k-1}\prod_{j=0}^i(N-j)$ are chosen such that fully separable states yield a vanishing value
of $T_k$. The advantage of this choice of entanglement monotone as compared to the various alternative options is that
the present objects can easily be evaluated in terms of the four point correlation functions that can be expressed in
terms of T\"{o}plitz determinants. $T_1$ describes the average entanglement between an individual spin and the rest of
the chain. $T_2$ includes in addition the entanglement contribution between pairs of spins and the residual system, and
$T_3$, and $T_4$ contain ever higher order correlations. This hierarchy extends to the multiparticle concurrence
\cite{andre} that provides a quantification for the entire entanglement content of a multipartite system. However, for
practical reasons we will restrict the present investigation to $T_1$-$T_4$ since the evaluation of $5$-point
quantities becomes impractical for large systems.

A clear difference between the behavior of bipartite and multipartite entanglement can be seen already going from $T_1$
to $T_4$ as shown in Fig.~\ref{fig1}, where the different multi-partite tangles are plotted as function of $\lambda$
for $\gamma=1$, and $\gamma=0.5$. In both cases neither $T_1$ nor $T_2$ show a distinct peak around the phase
transition, but their behavior rather resembles the step-like behavior of the bipartite concurrence \cite{osterloh}.
But starting from $T_3$ for the case of $\gamma=1$, and $T_4$ for $\gamma=0.5$, there is a clear peak arising around
the critical point. As it can be seen for $\gamma=1$ the peak becomes more pronounced with increased order of
correlation in $T_k$. This shows that the long-range correlations that are established around a quantum phase
transition are displayed in terms of multipartite entanglement, whereas the bipartite entanglement between blocks of
consecutive spins do not display this behavior.

\begin{figure}[htbp]
\begin{center}
\includegraphics[width=3.2 in]{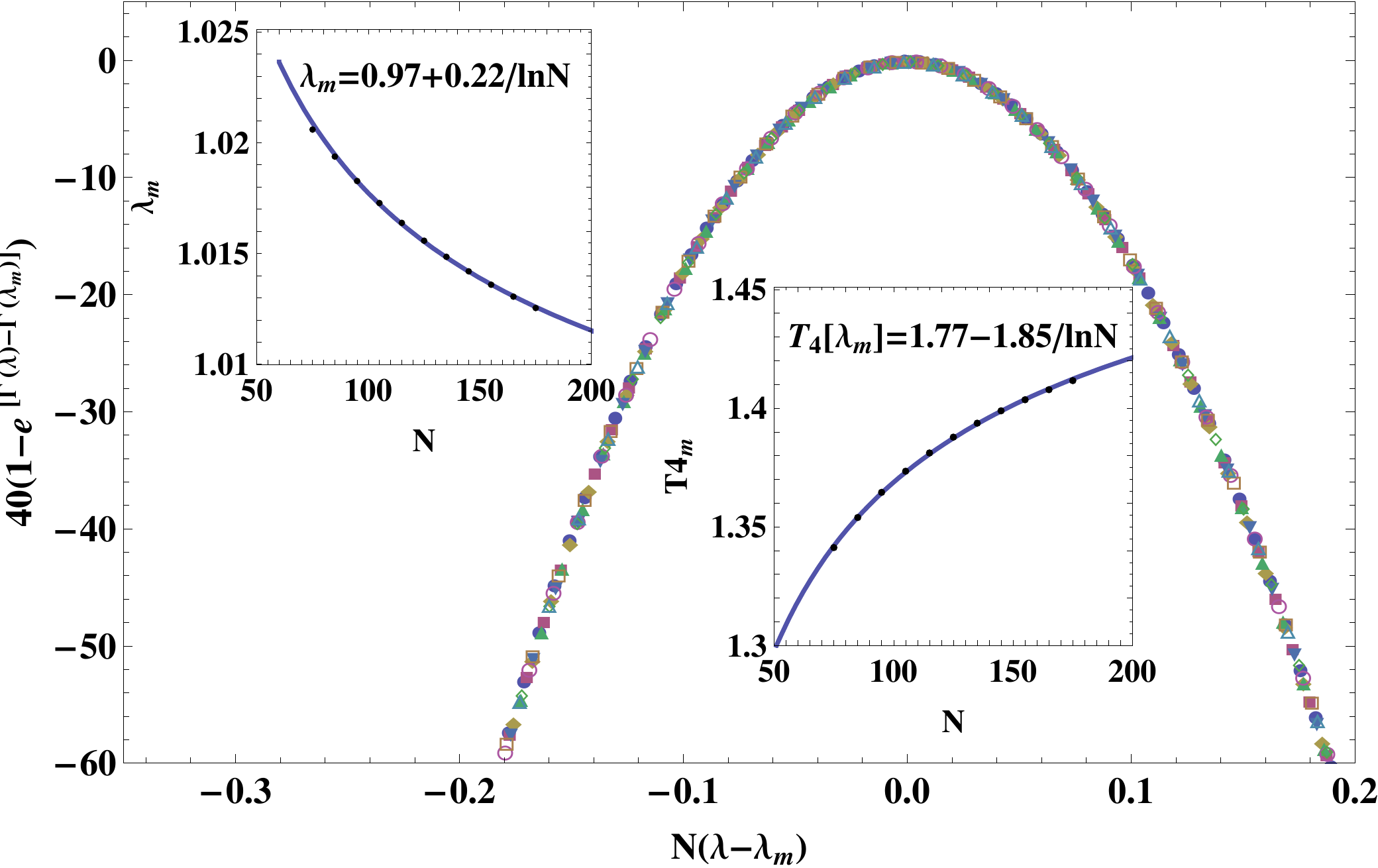}
\end{center}
\caption{The main plot shows the universality of  $\Gamma\equiv (T_4-T_4^*)^{-1}$ around the phase transition: $\Gamma
(\lambda)-\Gamma (\lambda_m)$ is plotted for various system sizes, from $N=75$ to $175$, as function of
$N(\lambda-\lambda_m)$, and the same behavior is found for all system sizes that are displayed with different symbols.
The insets show the growth (right) and position (left) of the peak with increasing system size.} \label{fig2}
\end{figure}

In order to quantify the growth of multipartite entanglement in $T_4$ at the critical point we study the  scaling with
system size of its peak height  and position. The right inset of Fig.~\ref{fig2} shows the growth of the peak with
increasing  number of spins. Since $T_4$ is a bounded quantity, it can not diverge even in the thermodynamic limit.
Nevertheless, the increase follows a logarithmic fit (solid line). The left inset of Fig.~\ref{fig2} shows that, with
increasing system size,  the position of the peak, $\lambda_m$,   gets shifted towards the  expected value  in the
thermodynamic limit, $\lambda_m=1$, however, according to the logarithmic fit that follows the data, it  does not reach
$\lambda_m=1$. Instead it overshoots a bit towards the value of $0.97$. We do not attribute too much meaning to this
observation, but rather expect that more data for larger systems would result in a slightly more accurate
extrapolation.

Considering the well known universal behavior of Ising-like models at a phase transition, it is central to check
whether our presently utilized quantification of entanglement reveals universality. Since by  definition $T_4$ is
always finite  a more appropriate quantity to study universal scaling is $\Gamma \equiv (T_4-T_4^*)^{-1}$ since it
exhibits a logarithmic divergence at the critical point. Here $T_4^*$ is the maximum value reached by $T_4$ in the
thermodynamic limit at $\lambda_m=1$, which we calculate to be $\pi/2$. Again $T_4^*$ does not agree exactly with the
value we get from our finite size scaling which yields 1.77 however as we mention before we attribute it  to the size
limitation we have in calculating four point correlations and expect a more  accurate fit for larger systems.
 Fig.~\ref{fig2} shows that $\Gamma(\lambda)-\Gamma(\lambda_m)=f[N^{1/\nu} (\lambda-\lambda_m)]$
exhibits universal behavior  with the scaling exponent $\nu=1$  characteristic for the Ising model \cite{Sachdev}: When
we plot this quantity  as function of $N(\lambda-\lambda_m)$ for  different system sizes, there is an almost  perfect
coincidence of the various data points into a single curve. Furthermore, the shape of the universal curve closely
resembles the universal curve underlying concurrence \cite{osterloh}.

\begin{figure}[htbp]
\begin{center}
\includegraphics[width=0.4\textwidth,angle=0]{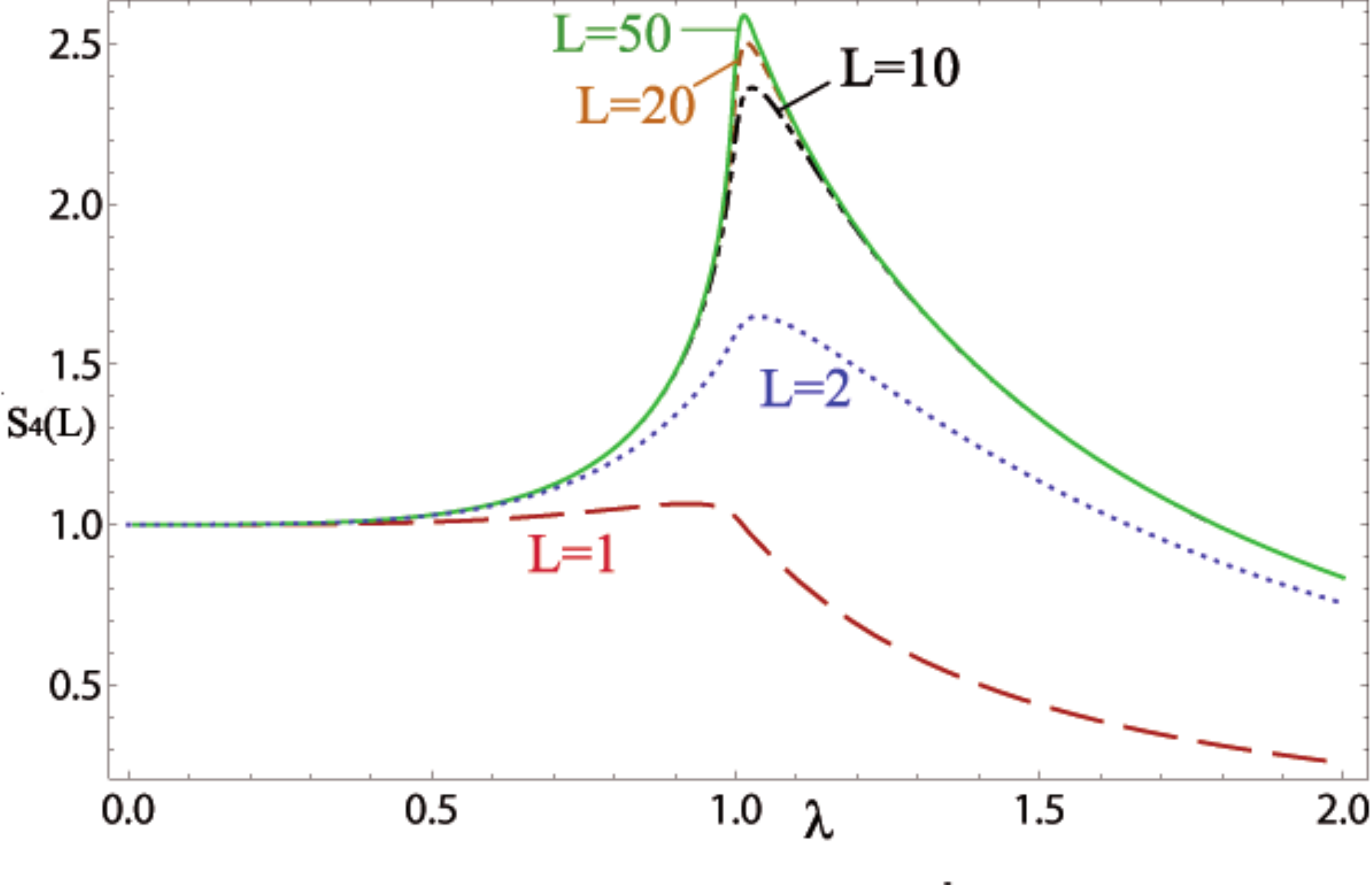} \leavevmode
\end{center}
\caption{Entanglement entropy, $S_4(L)$,  as function of $\lambda$ for different  spacings between the four spins.
$S_4(L)$  grows monotonously with increasing separation and reaches  its maximum value when $L=N/4$. Here $N=201$.}
\label{fig4}
\end{figure}

In order to show that it is exactly the long range correlations that grow, as opposed the entanglement between close
neighbors, and that the above findings are not a particularity of the employed tangles $T_i$, we present a similar
analysis as above also for a different, well established entanglement measure, the {\it entanglement entropy}. In
contrast to previous studies \cite{Latorre,Vidal} which considered bipartition of the spin chain into $\ell$
consecutive spins and the rest of the chain  here we study the entanglement entropy $S_4(L)$ extracted by  bipartition
of the spin-chain into four spins that are separated by equal distance $L$, (say spin $1$, $1+L$, $1+2L$, and $1+3L$)
and the rest of the system. This quantity is shown in  Fig.~\ref{fig4}. Note that similarly to $T_4$, also $S_4(L)$
remains finite in the thermodynamic limit, $S_4(L)\leq 4$. While there is only a tiny enhancement of the entanglement
entropy around the phase transition for $L=1$, a peak grows with increasing separation $L$ of the spins, and takes its
maximum for $L\simeq N/4$, {\it i.e.} in the case of maximum separation. This behavior  again gives evidence of the
long-range character of the correlations.

Having defined proper entanglement quantifiers that capture the properties of a quantum phase transition we will now
address the issue of how the latter can be linked to experimentally accessible  observables. Here we will focus our
discussion on  cold atomic systems in view of  their appeal  as ideal quantum simulators of iconic condensed matter
Hamiltonians. In these systems absorption  images taken after releasing the atoms from the trap are  the commonly used
diagnostic tools:
 the average density profile after time of flight  maps to   the quasi-momentum distribution $n(q)$ of the atoms at the
 release time, and
 the density-density correlations known as {\it noise correlations} \cite{Altman} yield the momentum-momentum
 correlations $\Delta(q_1,q_2)$ .
These functions are just Fourier transforms of two and four-point correlations of the atomic field creation and
annihilation operators at the various lattice sites, $\hat{a}_n, \hat{a}_n^\dagger$ :
\begin{eqnarray}
n(q)&\equiv& \langle \hat{n}_q\rangle=\frac{1}{N}\sum_{n,m} e ^{i\frac{2\pi}{N} q (n-m) } \langle
\hat{a}_n^\dagger\hat{a}_m\rangle\\ \Delta(q_1,q_2) &\equiv&  \langle (\hat{n}_{q_1}-\langle\hat{n}_{q_1}\rangle)(
\hat{n}_{q_1}- \langle\hat{n}_{q_1}\rangle)\rangle )\ . \label{noise}
\end{eqnarray}

The noise correlations $\Delta(q_1,q_2)$ contain the same non-local correlations as those included in $T_4$ and
$S_4(L)$. In contrast to $T_4$ and $S_4$, $\Delta(q_1,q_2)$ is not a proper entanglement monotone, but it is accessible
to measurement. Therefore, the natural question that we address below is: how much information about the enhancement of
many-body entanglement at the critical point can be inferred from noise correlations.

In view of the well known mapping between hard-core bosons (HCB) and spin-$1/2$ operators \cite{Lieb}, noise
correlations for the spin chains can be defined by using the following relations between the spins and atomic
operators: $\sigma^+_j=\hat{a}_j^\dagger$,   $\sigma^-_j=\hat{a}_j$ and $\sigma^z_j=\hat{a}_j^\dagger \hat{a}_j $. In
the following we use this mapping to study the behavior of   $\Delta(0,0)$ and $n(0)$ as $\lambda$ is varied across the
critical point. Our aim is to use this exactly solvable system  as a   benchmark of the behavior of these correlations
in  more general systems, which  are harder to deal theoretically but on the other hand where noise correlations   can
be accessed experimentally. Whereas their quantitative properties  will  depend on  the details of the underlying
Hamiltonian,  we believe their general features   will be generic  due to the  inherent universal scaling  of a quantum
phase transition.

\begin{figure}[htbp]
\begin{center}
\includegraphics[width=0.4\textwidth,angle=0]{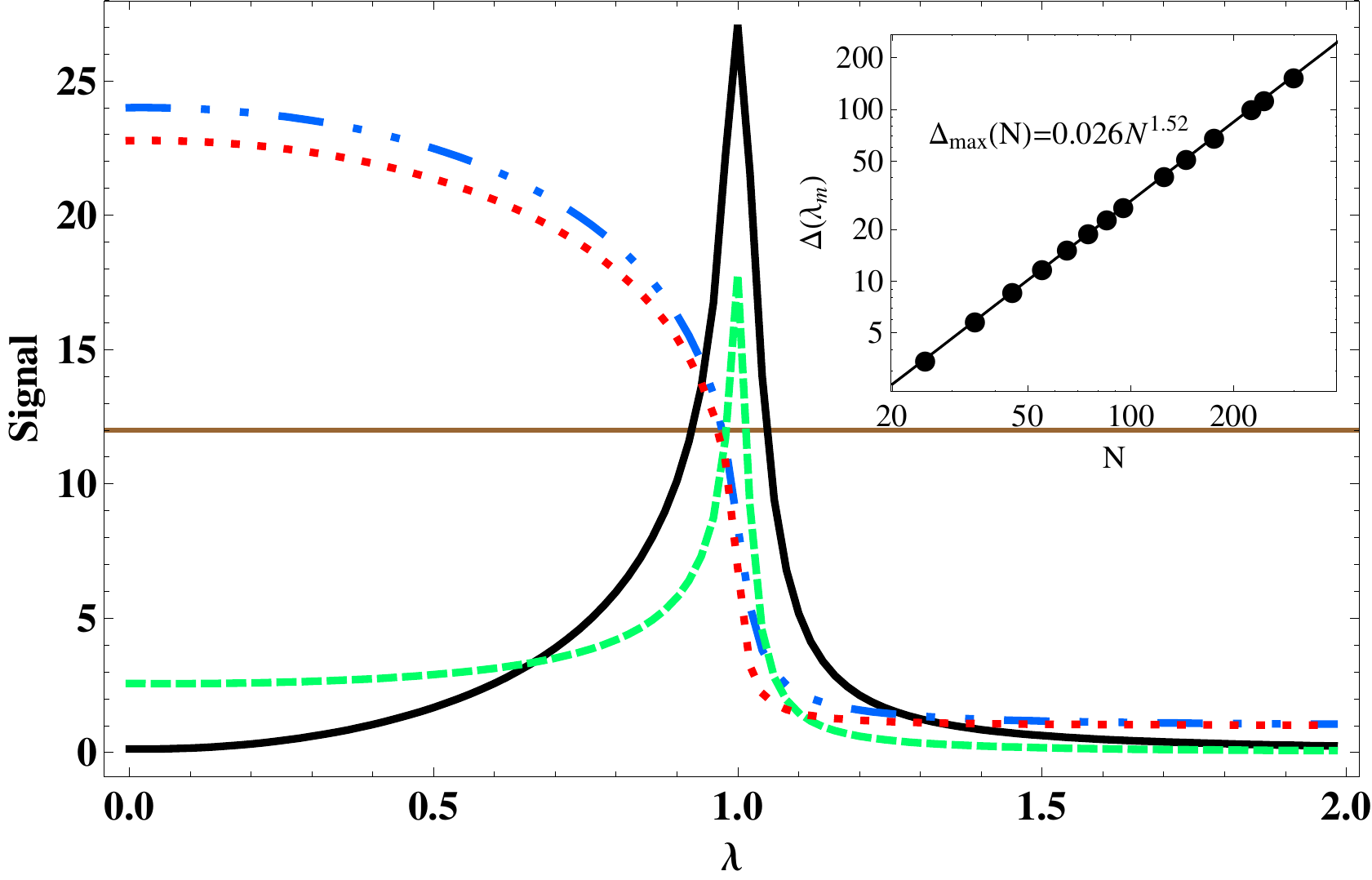} \leavevmode
\end{center}
\caption{(color online) $n(0)$ and  $\Delta(0,0)$ as a function of $\lambda$ for $N=95$ spins and two different
anisotropies: The dot-dashed blue (red dotted) curves and solid black (dashed green) correspond to n(0) and
$\Delta(0,0)$  for $\gamma=1$($\gamma=0.5$). The horizontal line at $12$ is the maximal value that $\Delta(0,0)$ can
take for a separable state and the peaks of $\Delta(0,0)$ exceed this threshold. The inset shows the numerical scaling
of the $\Delta(0,0)$ peak vs  $N$ which is in agreement with the analytic prediction of a powerlaw with exponent
$3/2$.} \label{fig5}
\end{figure}

Fig.~\ref{fig5}  shows that $\Delta(0,0)$ captures the enhancement of entanglement at the quantum critical point,
whereas $n(0)$ fails to do so. While $n(0)$ exhibits a step-like functional dependence of $\lambda$ (similar to $T_2$),
the noise autocorrelation function $\Delta(0,0)$ is sharply peaked around the phase transition (similar to $T_4$).
$\Delta(0,0)$ contains the higher order non-local  correlations which  are the key to properly extract the enhancement
of entanglement at the quantum phase transition.

Our calculations for noise correlations also verify that they can be related to entanglement because they probe
separability. For any separable state $\ket{\Phi_s}$ of a spin-$1/2$ system ({\it i.e.} a state that can be expressed
in terms of one-body states $\ket{\phi_i}$ as
$\ket{\Phi_s}=\ket{\phi_1}\otimes\ket{\phi_2}\otimes\hdots\otimes\ket{\phi_N}$), yields a value of $\Delta(0,0)$ that
is smaller than \cite{inpreparation} \be \Delta_{\mbox{max}}(\Phi_s)=\frac{1}{8}\left(1+N \right)\label{separability}
\ee That is, if the obtained value of $\Delta(0,0)$ exceeds this threshold then the underlying state it entangled. In
Fig.~\ref{fig5} we indicate $\Delta_{\mbox{max}}(\rho_s)$  for  $N=95$ . The peak of $\Delta(0,0)$ clearly exceeds the
separability threshold and  the excess becomes even more pronounced with increasing system size. The reason is the
following: While  $\Delta(0,0)$  grows algebraically with the system size as $N^{3/2}$, the separability threshold of
Eq.~\eqref{separability} grows linearly with $N$, so that the excess of the peak over this threshold increases as
$\sqrt{N}$. The exponent $3/2$ in the noise correlation peak height follows from the universal scaling  of $\langle
\sigma_x\rangle \sim (\lambda-1)^{1/8}$ which yields a divergence  of the noise correlation peak as
$(\lambda-1)^{-3/2}$ in the thermodynamic limit and to a $N^{3/2}$ scaling for finite samples.

Even though the above investigations were focused on the anisotropic XY model Hamiltonian, we emphasize that  Eq. (5)
provides a general criterion for arbitrary spin systems and consequently corresponds to a useful benchmark  to study
entanglement properties in more general  systems such  disordered spin or spins at  finite temperature. Additionally,
preliminary calculations done in in comparatively small
  1D soft core bosons undergoing a  superfluid to Mott insulator transition \cite{Rey, RIC} also
  indicate a similar peaked behavior  of $\Delta$ close to the critical point. This  suggests  that
   the growth of $\Delta$
   can be a generic signature of a quantum phase transition  and that    noise correlations are suitable
   observables to experimentally verify the enhancement of entanglement.



F.M. acknowledges  financial support by Alexander v. Humboldt foundation and the ITAMP visitors programm, and A.M.R  by
NSF through a grand from ITAMP.

\end{document}